\documentclass[11pt]{article}

\usepackage[margin=1in]{geometry}
\usepackage{graphicx}
\usepackage{booktabs}
\usepackage{microtype}
\usepackage{xspace}
\usepackage{enumitem}
\usepackage{hyperref}
\usepackage{placeins}

\hypersetup{
  colorlinks=true,
  linkcolor=black,
  citecolor=black,
  urlcolor=blue
}

\newcommand{\system}{\textit{The Architect}\xspace}

\title{\system: Interactive Visualization of Deep Learning Mathematics Directly in Microsoft Excel\\[0.6em]Technical Report}
\author{Mohammad Imrul Jubair \quad Tom Yeh\\University of Colorado Boulder}
\date{July 1, 2026}

\begin{document}

\maketitle

\begin{abstract}
We present \system, a system that turns Microsoft Excel into an interactive view of deep learning mathematics. A user describes a neural network in a compact table. The system then generates a workbook that shows the full forward pass and, when requested, the backward pass and parameter updates. Computed values appear as live spreadsheet formulas, while user-controlled values such as inputs, weights, labels, and hyperparameters remain editable. Excel reactively updates the dependent computations through its recalculation engine.

Most deep learning tools hide the numerical details behind library calls. Many visualization tools show architecture diagrams or training summaries, but they do not expose the full arithmetic of the model. \system focuses on that missing middle layer. It makes matrices, activations, losses, gradients, and updates visible as inspectable spreadsheet regions, with editable controls for values users naturally manipulate. The system also produces aligned PyTorch snippets, which helps users connect formulas to implementation.

This report describes the motivation, design, implementation, and use cases of \system. We show how the system supports introductory arithmetic tracing, learning-rate exploration, diagnosis of dying ReLU, and inspection of vanishing gradients. The main idea is simple: spreadsheets already support formulas, direct editing, reactive recomputation, and tabular layout. These properties make them a useful medium for understanding how small educational and diagnostic neural networks compute.
\end{abstract}

\tableofcontents

\section{Introduction}
Deep learning is often taught and used through software that hides most of the arithmetic. A learner can call \texttt{loss.backward()} without seeing how the gradients are formed. A practitioner can tune a model without directly seeing where gradients shrink, explode, or disappear. This is useful for productivity, but it creates a gap between the code people run and the computations the model performs.

Existing visualization tools help in many ways, but they usually focus on model structure, data flow, or output behavior. They less commonly show the full matrix arithmetic of a forward pass, and backward-pass views are often summarized rather than exposed as inspectable, cell-level computations.

This report explores a different approach. Instead of building a new visualization environment, we use a familiar spreadsheet interface. \system uses Microsoft Excel as a live canvas for neural network mathematics. A user specifies a network in a small table. The generator then creates a workbook that lays out inputs, weights, activations, losses, gradients, and updates as a mixture of editable controls and native spreadsheet formulas.

Because the workbook uses live formulas, it stays reactive after generation. If a user changes a weight, input, target label, or learning rate, Excel updates dependent values through recalculation. This makes the workbook useful not only for explanation, but also for debugging and experimentation.

The main contributions of this work are:
\begin{enumerate}[leftmargin=*]
\item A system that generates reactive spreadsheet blueprints for neural network computation from a compact architecture specification.
\item A spreadsheet-native interaction style for inspecting and editing deep learning arithmetic directly.
\item A set of illustrative scenarios that show how the representation supports learning and diagnosis.
\end{enumerate}

We frame these contributions as a systems and interaction design contribution rather than a controlled learning-outcomes claim. The report supports the design through illustrative scenarios and classroom deployment observations; formal evidence of improved learning, transfer, usability, or debugging performance remains future work.

\begin{figure}[htbp]
  \centering
  \includegraphics[width=\textwidth]{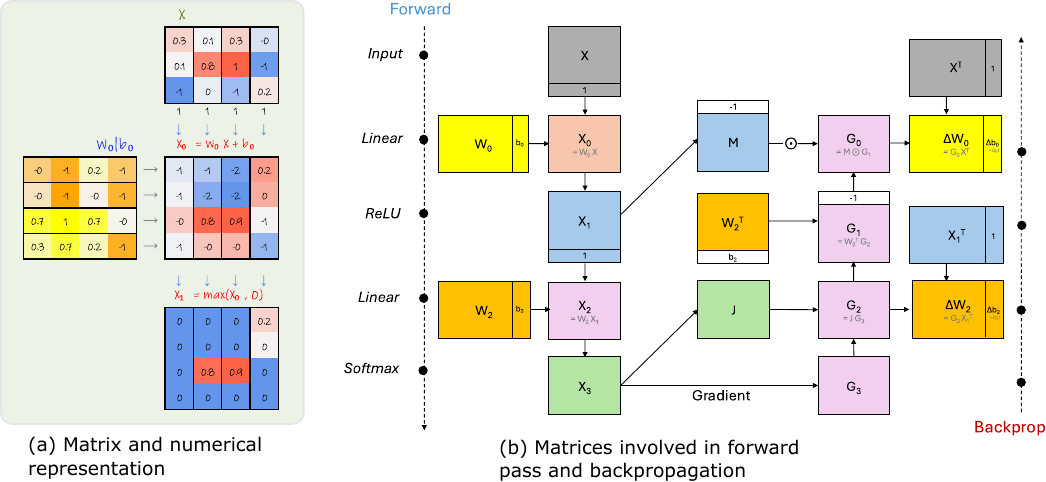}
  \caption{Matrix-first workflow. \system treats forward and backward propagation as explicit matrix workflows and then renders them as live spreadsheet formulas.}
  \label{fig:matrix-principle}
\end{figure}

\begin{figure*}[htbp]
  \centering
  \includegraphics[width=\textwidth]{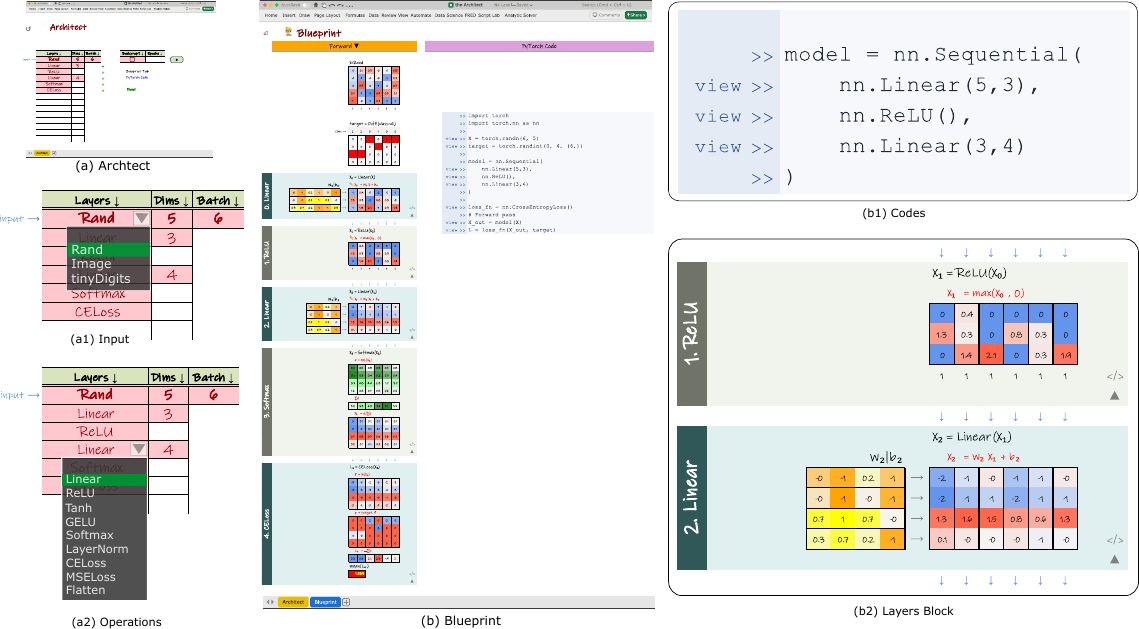}
  \caption{System overview. A user specifies a network in the Architect tab and runs the generator. The workbook then produces a live mathematical blueprint in Excel. Editing a weight, input, or hyperparameter updates dependent cells through Excel's calculation engine.}
  \label{fig:teaser}
\end{figure*}

\FloatBarrier

\section{Background and Motivation}
\subsection{Why Current Tools Leave a Gap}
Modern deep learning frameworks are excellent execution engines. They are not designed to make every intermediate value easy to inspect. This is a good trade-off for large-scale training, but it is a weak fit for learning and debugging when people need to see the numerical process itself.

Visualization tools improve access, but most of them work at a higher level of abstraction. Systems such as TensorFlow Playground, CNN Explainer, Transformer Explainer, Netron, and VisualKeras help users understand model behavior or structure in different ways~\cite{smilkov2017tensorflow,wang2020cnn,hong2024transformer,roeder2023netron,gavrikov2021visualkeras}. These views are useful, but they do not usually reveal the full chain of matrix operations, derivatives, and updates.

This gap matters in both teaching and practice. In teaching, students often learn equations on slides and then jump directly into framework code. In practice, engineers often observe training symptoms such as unstable loss or weak gradients without seeing the exact arithmetic path that caused them. In both settings, the missing layer is the concrete numerical state of the model.

\subsection{Why a Spreadsheet Is a Good Fit}
Spreadsheets already provide several properties that match this problem well. Prior work has long treated spreadsheets as a meaningful end-user programming environment and as a substrate for richer interactive systems~\cite{nardi1990spreadsheet,kandel2011wrangler,smith2009nodexl,chang2014interactiveweb}. Spreadsheet-based learning tools have also been used for machine-learning and neural-network education~\cite{thinYin2020simpler,semerikov2020spreadsheets,roshanaei2024logistic}. In our case, the fit is especially strong because neural network arithmetic is naturally tabular.

Spreadsheets provide several useful properties:
\begin{itemize}[leftmargin=*]
\item They organize information in rows and columns, which maps naturally to vectors and matrices.
\item They expose formulas directly through the formula bar and cell references.
\item They support direct editing of values.
\item They recalculate dependent cells automatically.
\item They are familiar to many users.
\end{itemize}

These are ordinary spreadsheet features, but in this project they become mechanisms for understanding model arithmetic.

Another reason the spreadsheet works is pacing. Users can inspect one cell, one row, or one layer at a time. They do not need to understand the entire model at once. This supports a gradual style of exploration that fits both learning and debugging.

\subsection{Transparency Gap}
Three problems motivate the system.

\textbf{Black-box interiors.} Standard tooling hides the numerical middle of the model. Users can often see inputs and outputs, but not the full path in between.

\textbf{Interface friction.} Many explanatory systems require users to learn a new interface at the same time they are trying to learn the mathematics. A spreadsheet lowers that cost.

\textbf{Backpropagation blind spots.} Existing tools often emphasize the forward pass or model structure. Training is usually summarized through scalar metrics rather than explicit gradient pathways.

These three issues point to the same opportunity. If a system can expose the internal arithmetic in a familiar environment, then many important concepts become easier to inspect directly instead of being inferred from code or summarized plots.

\FloatBarrier

\section{Design Goals}
We designed \system around four goals.

\textbf{G1: Mathematical transparency.} Users should see real numbers and formulas, not just abstract diagrams.

\textbf{G2: Training completeness.} The system should support both inference and learning, including losses, gradients, and updates.

\textbf{G3: Low interaction overhead.} The interface should build on a familiar environment instead of a new custom tool.

\textbf{G4: Reactive manipulability.} Users should be able to change important values and see the effects after workbook recalculation.

\FloatBarrier

\section{System Overview}
\subsection{Workflow}
\system is distributed as an Excel workbook plus an Office Script. The workbook contains the specification table, helper sheets, and example data. The script reads the specification and generates the visualization sheets. The project repository is available as a \href{https://github.com/imruljubair/The-Architect}{GitHub repository}. A \href{https://docs.google.com/spreadsheets/d/1YeqPpTiOqDD-Xdamez5HHDONuGBbzATqW03fZR_N5-c/edit?usp=sharing}{Google Sheets version} is also available. The Google Sheets version is slower than running \system locally in Microsoft Excel, but it is useful for readers who want to avoid installing or using Excel. We also provide a \href{https://youtu.be/HzAudh_PK8I?si=8r9r7UDQP2NpzCVI}{short tutorial video} showing how to use \system.

The user starts in the \emph{Architect} tab. Each row describes one layer or operation. The system currently supports random inputs, image-like inputs, and a bundled \texttt{tinyDigits} dataset. Supported operations include \texttt{Linear}, \texttt{ReLU}, \texttt{Tanh}, \texttt{GELU}, \texttt{LayerNorm}, \texttt{Flatten}, \texttt{Softmax}, cross-entropy loss, and mean-squared-error loss.

The user can generate either:
\begin{itemize}[leftmargin=*]
\item an inference-only blueprint, or
\item a training blueprint with backpropagation, multiple epochs, and a dashboard.
\end{itemize}

\begin{figure*}[htbp]
  \centering
  \includegraphics[width=\textwidth]{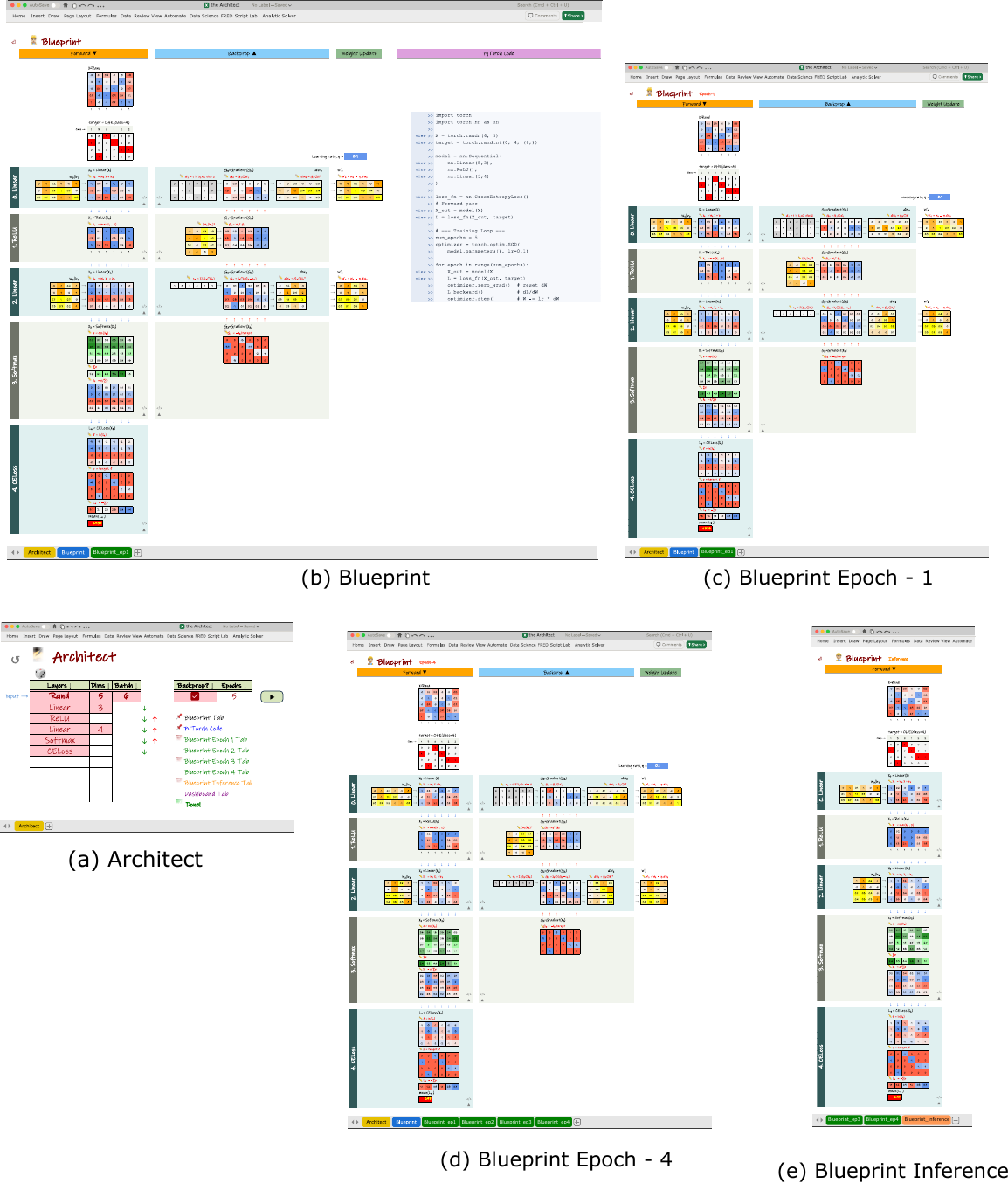}
  \caption{Workbook organization with backpropagation enabled. Persistent tabs provide the specification surface, helper data, and base blueprint. Generated epoch sheets and a dashboard extend the workbook into a multi-epoch training trace.}
  \label{fig:workflow}
\end{figure*}

\FloatBarrier

\subsection{Specification Interface}
The Architect tab is intentionally simple. Each row specifies a layer type and, when needed, an output size. The first row also specifies the input type and batch size. Extra controls enable backpropagation and choose the number of epochs to materialize.

This simple interface matters. Users can create many network variants quickly, but the structure is still regular enough for the script to generate the correct formulas and references.

The low authoring burden is important. A user can change one row, regenerate the workbook, and compare a new variant without rewriting model code. This makes the system feel closer to sketching than to programming.

\begin{figure}[htbp]
  \centering
  \includegraphics[width=\textwidth]{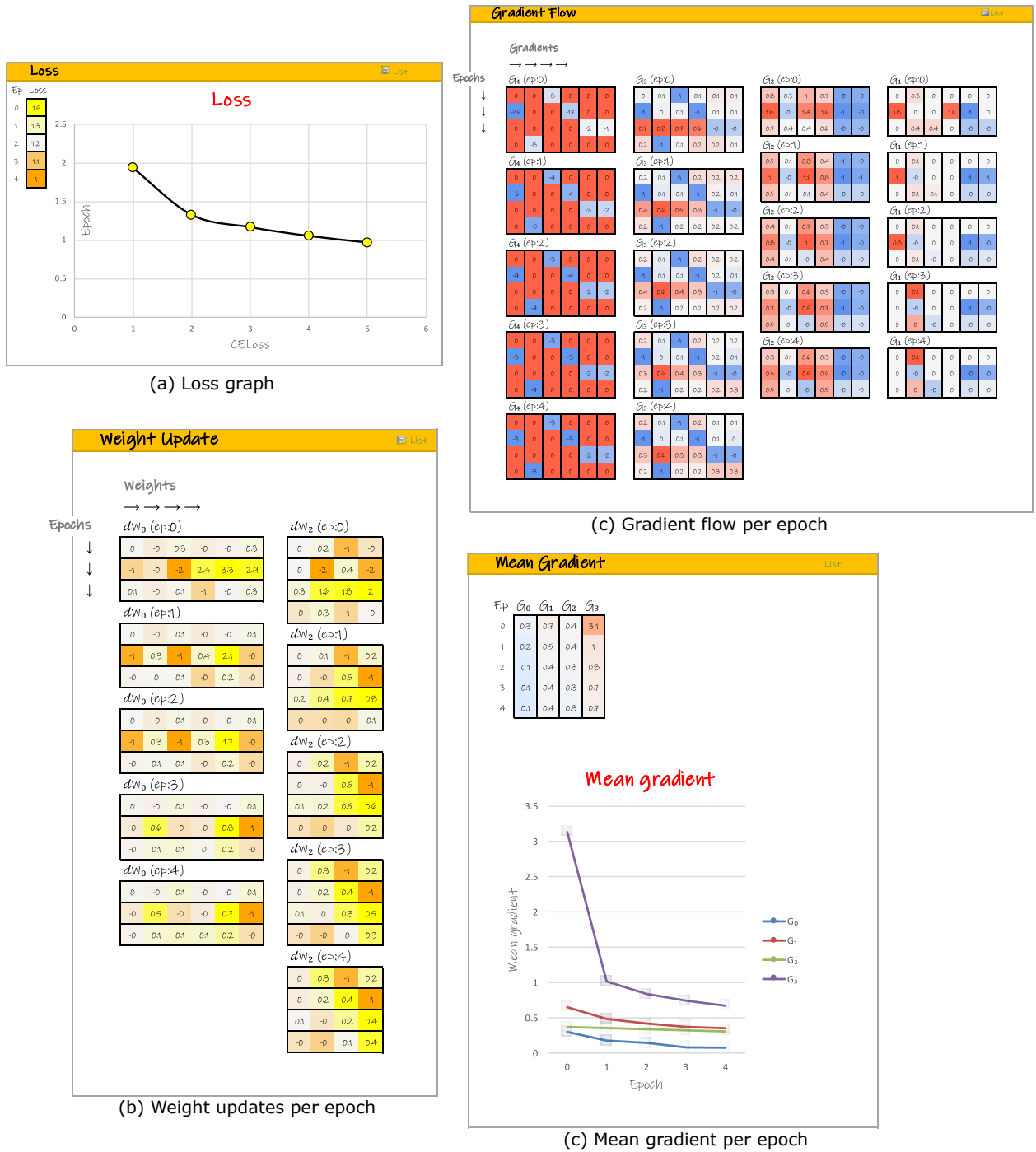}
  \caption{Specification interface in the Architect tab. The user declares the network one row at a time and can choose whether to generate backpropagation and epoch sheets.}
  \label{fig:architect}
\end{figure}

\FloatBarrier

\subsection{Blueprint Representation}
The core output is the \emph{Blueprint} sheet. It uses a left-to-right layout of operation blocks. Each block shows a layer's inputs, parameters, and outputs. Linear layers show matrices for the input, weights, bias terms, and outputs. Nonlinearities show the element-wise transformation and its result.

This representation works because spreadsheet grids and matrix arithmetic fit each other well. The workbook does not just describe the computation. It contains the computation in visible form.

Each block also acts like a local explanation. Users can understand one layer as a small transformation before moving to the next one. The result is easier to follow than a large sheet of unlabeled values.

\begin{figure}[htbp]
  \centering
  \includegraphics[width=\textwidth]{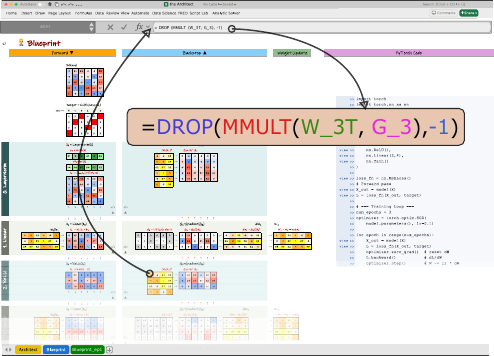}
  \caption{A generated blueprint block. Each operation appears as labeled matrices and formula cells, so intermediate computations stay inspectable while user-controlled values remain editable.}
  \label{fig:blueprint}
\end{figure}

\FloatBarrier

\subsection{Backpropagation and Multi-Epoch Views}
When backpropagation is enabled, \system adds losses, local derivatives, incoming gradients, parameter gradients, and update equations. It also creates separate sheets for later epochs. These sheets are linked by formulas, so the updated parameters from one epoch feed into the next.

This design makes training persistent. Users can inspect a specific epoch as a stable object instead of watching a transient loop that disappears after execution.

That persistence is especially useful for classroom discussion and debugging. An instructor or collaborator can point to a single epoch sheet and talk about a specific weight change or gradient pattern without losing the surrounding context.

\begin{figure*}[htbp]
  \centering
  \includegraphics[width=\textwidth]{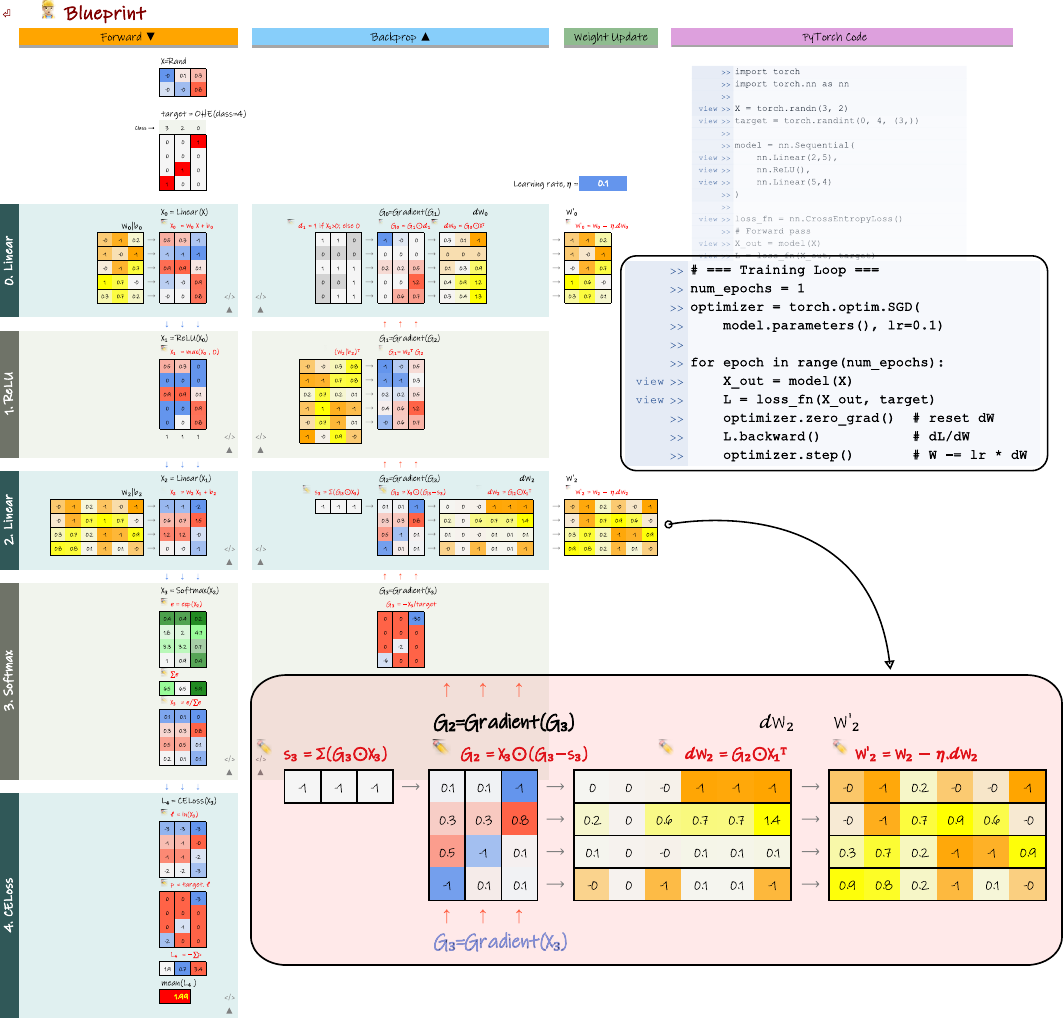}
  \caption{Detailed backpropagation view. The blueprint exposes losses, local derivatives, incoming gradients, and parameter updates as explicit spreadsheet regions.}
  \label{fig:backpropdetail}
\end{figure*}

\begin{figure}[htbp]
  \centering
  \includegraphics[width=\textwidth]{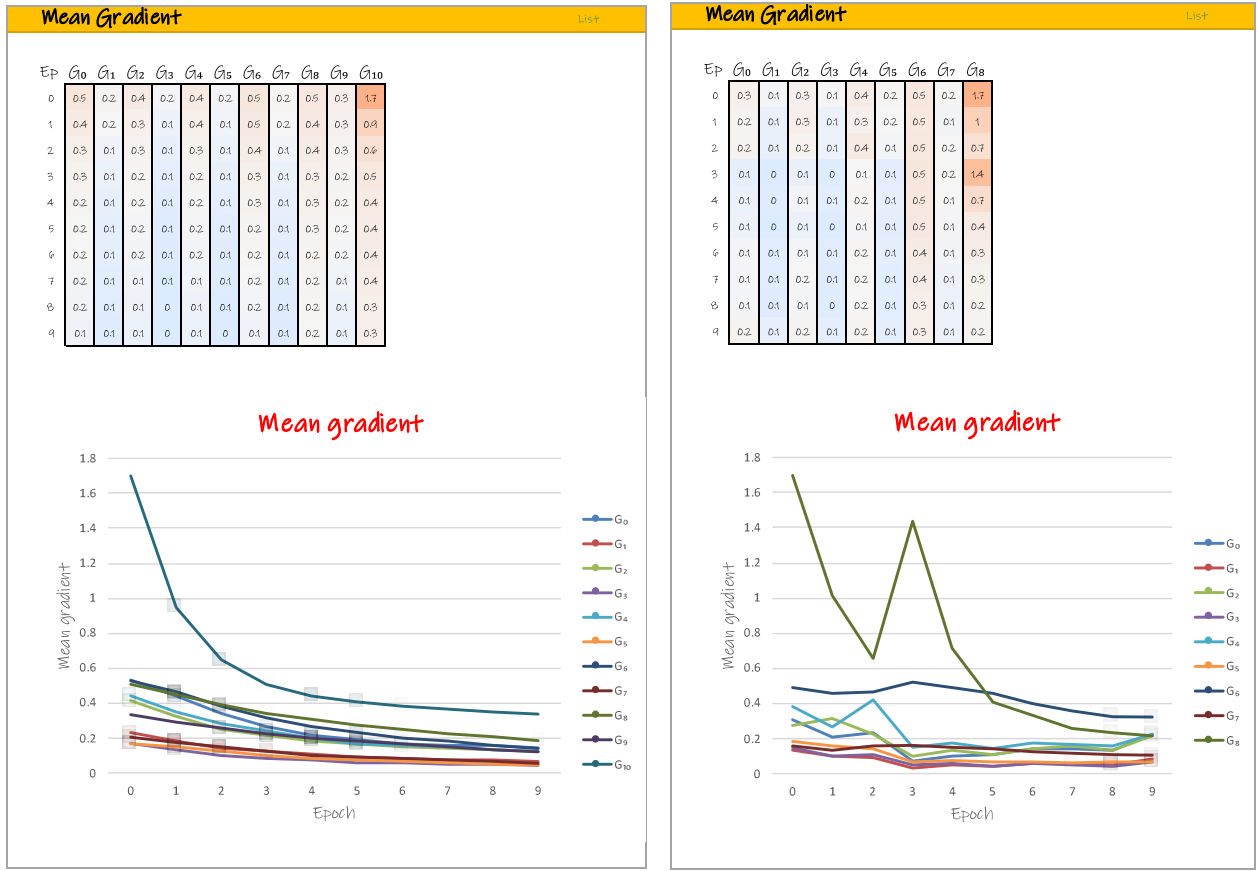}
  \caption{Dashboard view for training behavior. The dashboard summarizes the loss and layer-wise changes across epochs while staying linked to the underlying epoch sheets.}
  \label{fig:dashboard}
\end{figure}

\FloatBarrier

\subsection{Data and Code Bridges}
The workbook includes a built-in \texttt{tinyDigits} dataset so users can run classification examples directly in Excel. The system also generates aligned PyTorch snippets next to the spreadsheet blueprint. This helps users move between the formula-level view and the code-level view of the same model.

\begin{figure}[htbp]
  \centering
  \includegraphics[width=\textwidth]{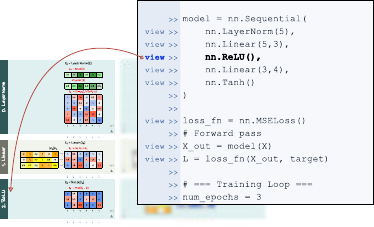}
  \caption{Linked code blocks. The workbook emits PyTorch snippets alongside blueprint regions and links code fragments back to the corresponding layer blocks.}
  \label{fig:codeblocks}
\end{figure}

\begin{figure}[htbp]
  \centering
  \includegraphics[width=\textwidth]{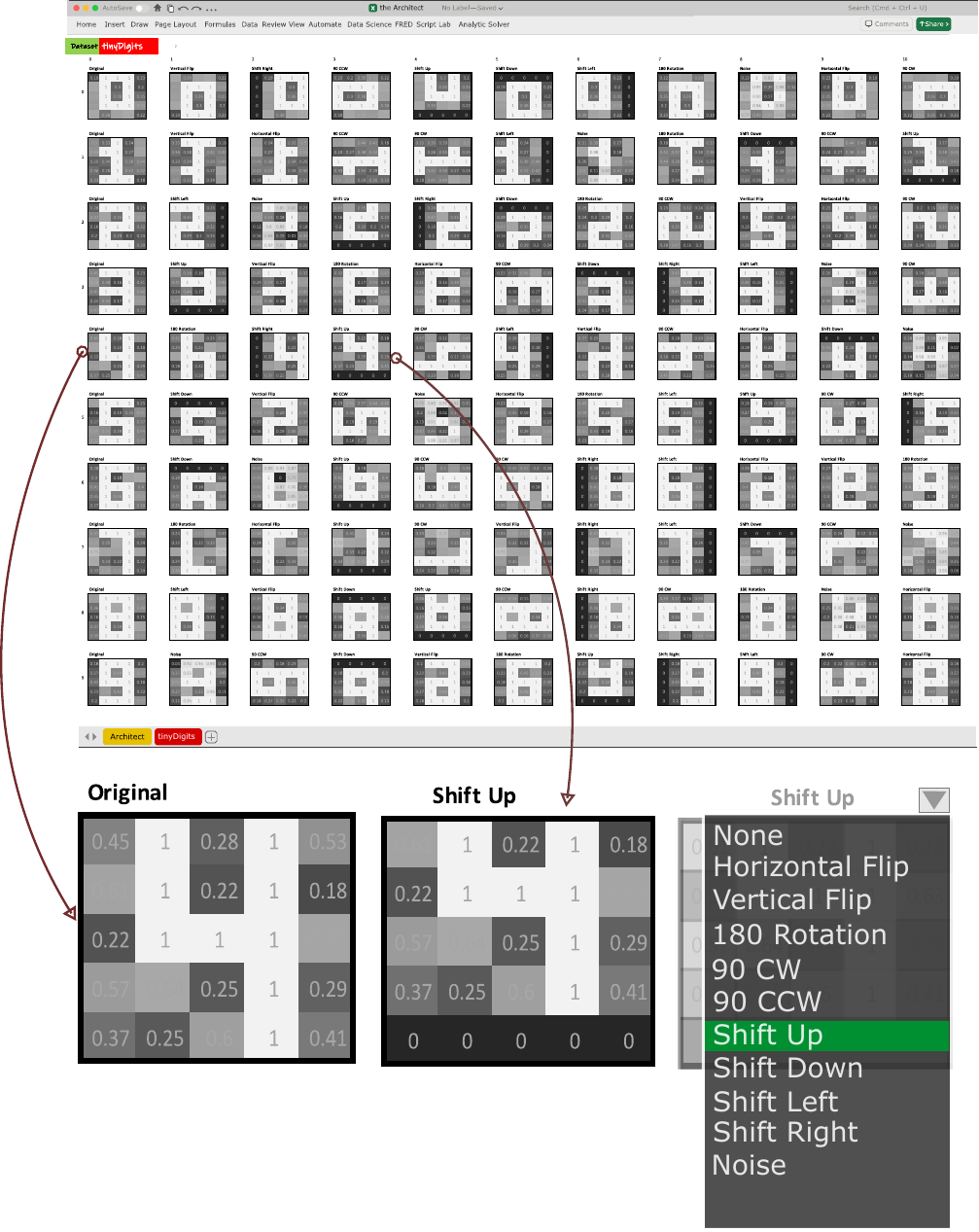}
  \caption{The built-in \texttt{tinyDigits} dataset. Each sample is represented as a small pixel grid, which supports direct classification examples inside the workbook.}
  \label{fig:dataset}
\end{figure}

\FloatBarrier

\subsection{Spreadsheet-Native Interaction Techniques}
Much of the system's value comes from features that already exist in Excel:
\begin{itemize}[leftmargin=*]
\item selecting a cell to inspect its formula,
\item using named ranges to navigate important variables,
\item using dynamic arrays to initialize blocks of values,
\item using cross-sheet references to chain epochs,
\item tracing precedents and dependents to inspect dependencies, and
\item duplicating sheets or workbooks for comparison.
\end{itemize}

\begin{figure}[htbp]
  \centering
  \includegraphics[width=\textwidth]{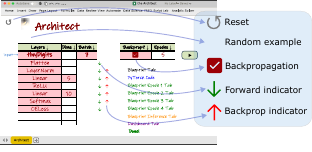}
  \caption{Additional workbook affordances. Native spreadsheet features support exploration, editing, and fast inspection of model state.}
  \label{fig:features}
\end{figure}

\begin{figure*}[htbp]
  \centering
  \includegraphics[width=0.76\textwidth]{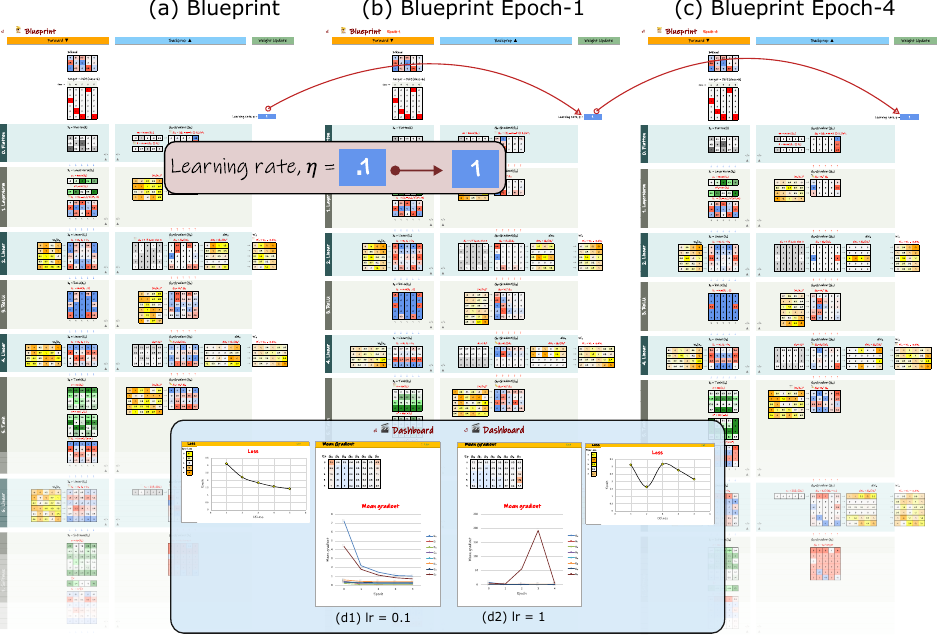}
  \caption{Spreadsheet-native affordances become model-inspection techniques. Editing cells such as class indices or hyperparameters triggers workbook-wide recomputation.}
  \label{fig:affordances}
\end{figure*}

\FloatBarrier

\section{Implementation}
\subsection{Office Script Generator}
The generator is implemented as an Office Script in TypeScript. It reads the Architect tab, infers the layer sequence and dimensions, clears previously generated content, and constructs new workbook content programmatically.

The script keeps a registry of ranges, variables, and named items. This registry is important because later formulas depend on addresses created earlier in the generation process.

\subsection{Formula Templates and Address Resolution}
The workbook stores formulas, not just values. This means each operation must be generated as a parameterized formula template with the correct cell addresses filled in at generation time.

This is straightforward for simple blocks, but it becomes harder when later regions depend on earlier regions and backward-pass formulas depend on forward-pass geometry. The registry helps manage this complexity by storing semantic information about each generated block.

\subsection{Extending the Generator}
The generator is intended to be extensible, but extension requires work at several levels. To add a new operation, a developer must register the operation name and its dimension rules, define the forward formula templates, add backward-pass templates if training support is desired, declare the named ranges that later formulas will reference, and specify how the operation should be laid out in the worksheet.

This workflow is direct for operations that share the same rectilinear matrix structure as the existing blocks. For example, another pointwise activation can reuse much of the existing activation layout. Operations such as convolution, attention, or residual connections are more involved because they require new layout primitives for spatial neighborhoods, multiple inputs, or non-sequential dependencies. These operations are therefore better understood as future extensions rather than simple additions to the current prototype.

\subsection{Dynamic Arrays and Reactivity}
The system relies on Excel's dynamic array engine. In many cases, the script writes a formula into an anchor cell and lets Excel spill the results into a surrounding matrix range.

This choice has two advantages:
\begin{itemize}[leftmargin=*]
\item displayed matrices stay inspectable through their formulas, and
\item user edits propagate through Excel's recalculation graph.
\end{itemize}

The trade-off is that generation logic becomes more complex because formulas must be assembled carefully.

This complexity is mostly hidden from the user. From the user's perspective, the important result is that the workbook remains live after generation. That liveness is central to the overall interaction style.

\subsection{Visual Structuring}
Showing real matrix arithmetic can easily become overwhelming. To reduce this problem, the workbook uses a consistent visual grammar:
\begin{itemize}[leftmargin=*]
\item forward-pass blocks flow from left to right,
\item gradient-related blocks flow back toward earlier layers,
\item repeated operations reuse the same layout conventions, and
\item labels, spacing, and grouping separate inputs, parameters, activations, losses, and updates.
\end{itemize}

These layout decisions do not remove complexity, but they make it easier to navigate. Repetition is helpful here. Once a user understands one linear block or one activation block, the same visual pattern appears again in later parts of the workbook.

\begin{figure}[htbp]
  \centering
  \includegraphics[width=\textwidth]{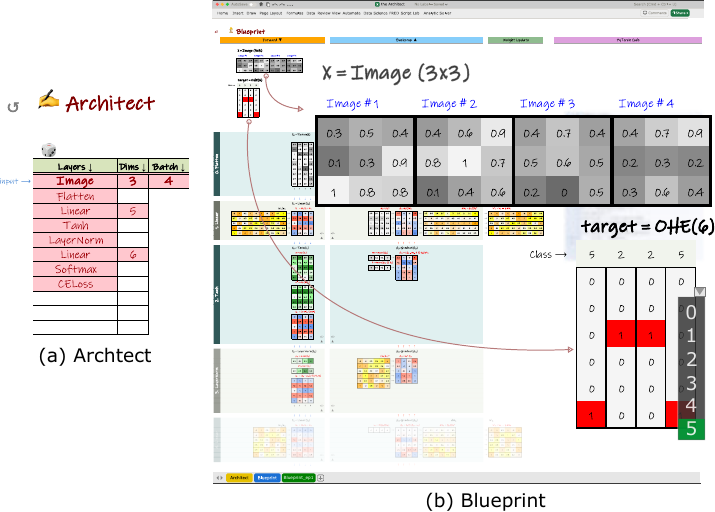}
  \caption{Image-shaped input support. The workbook can render an input as a two-dimensional pixel block, then flatten and route it through the same matrix-based blueprint.}
  \label{fig:imageinput}
\end{figure}

\FloatBarrier

\subsection{Supported Scope}
The current prototype focuses on small educational and diagnostic examples. It supports fully connected networks and a set of common operations that fit well into a rectilinear spreadsheet layout. This scope is intentional. The system is designed for inspectability, not large-scale training.

\FloatBarrier

\section{Using the System}
\subsection{Inference-Only Blueprints}
In inference mode, \system generates only the forward pass. This is useful for introductory explanation, shape reasoning, and inspection of intermediate activations without the added complexity of gradients.

\subsection{Training Blueprints}
When training mode is enabled, the workbook adds loss computation, backward regions, parameter updates, epoch sheets, and dashboard summaries. At that point, the workbook becomes a persistent training trace.

\begin{figure}[htbp]
  \centering
  \includegraphics[width=\textwidth]{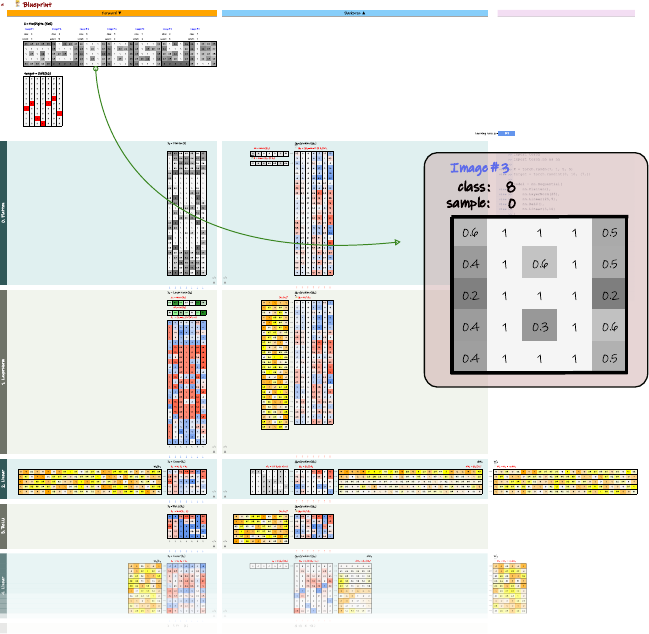}
  \caption{tinyDigits as direct model input. Users can select visible samples from the workbook dataset and feed them into the blueprint.}
  \label{fig:tinydigitsinput}
\end{figure}

\FloatBarrier

\subsection{Authoring and Comparison Workflow}
A typical workflow is simple:
\begin{enumerate}[leftmargin=*]
\item specify a network in the Architect tab,
\item generate the workbook blueprint,
\item inspect a behavior or phenomenon,
\item modify the table or workbook values, and
\item regenerate or compare variants.
\end{enumerate}

Because the layout is consistent across examples, users can compare architectures or hyperparameters by looking at corresponding regions.

This comparison workflow is one of the practical strengths of the system. A user does not need to mentally map between very different interfaces. The same workbook grammar is reused across examples.

\section{Illustrative Scenarios}
\subsection{Tracing Introductory Arithmetic}
For learners, \system makes it possible to inspect a small MLP one cell at a time. A user can click an output cell in a linear layer, read the formula, and follow the references back to a specific input and weight. If the user changes one weight, the downstream changes become visible after recalculation.

This supports a concrete ``AI by hand'' workflow. Instead of jumping from symbolic equations directly to framework code, users can work through actual numbers in a live environment. This direction is inspired in part by matrix-first explanatory materials such as By Hand and the Deep Learning Math Workbook~\cite{yeh2024byhand,yeh2025workbook}.

It also supports a slower and more careful style of inspection. Users can pause at any layer, verify a value, and only then continue. This is useful when the goal is understanding rather than speed.

The system has also been used in a graduate-level computer science course built around matrix-first deep learning instruction. In that setting, students used the Architect tab to generate workbook blueprints and submitted different blueprint versions for assignments.

\subsection{Classroom Deployment Observations}
We also have preliminary classroom deployment evidence. In one graduate AI course, 82 students received an Architect assignment and 76 submitted generated workbooks. Students used the Architect tab to specify architectures, generate blueprints, trace backward-pass computations, and submit resulting workbook artifacts.

These observations are not a controlled evaluation of learning outcomes. They do, however, provide evidence that students could use the system in an authentic course setting. The submitted workbooks also showed exploratory behavior. Many students stacked several linear operations in sequence to observe how values changed during backpropagation. Several students added their own Excel charts to analyze activations or gradients across layers. These additions suggest that the spreadsheet representation invited some students to extend the generated artifact rather than only view it passively.

\subsection{Diagnosing Dying ReLU}
The backward view makes activation-related failures easier to inspect. In networks with ReLU activations, users can see gradients become zero when the corresponding pre-activations are negative. If the user swaps ReLU for Tanh and regenerates the blueprint, the contrast appears directly in the gradient matrices.

\begin{figure}[htbp]
  \centering
  \includegraphics[width=\textwidth]{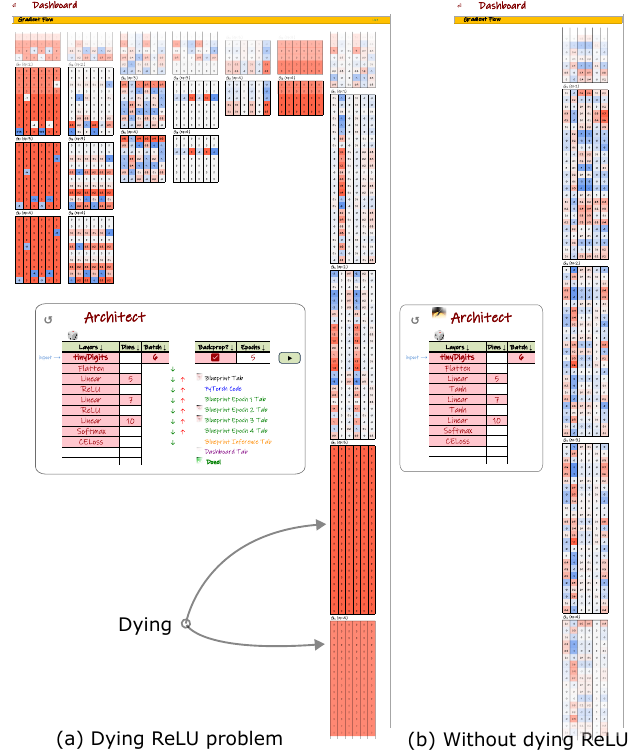}
  \caption{Dying ReLU and gradient diagnosis. The blueprint makes it possible to compare zeroed gradients under ReLU with recovered gradient flow under alternative activations.}
  \label{fig:dyingrelu}
\end{figure}

\subsection{Exploring Learning-Rate Sensitivity}
Because epoch sheets and dashboard summaries are formula-linked, changing one learning-rate cell affects update magnitudes, later weights, and summary behavior through workbook recalculation. This makes it easy to compare under-training, stable training, and divergence without rerunning an external script.

This reactive response is helpful because it shortens the feedback loop. Users can test a hypothesis with a small edit and see the consequences in both local formulas and aggregate summaries.

\begin{figure}[htbp]
  \centering
  \includegraphics[width=\textwidth]{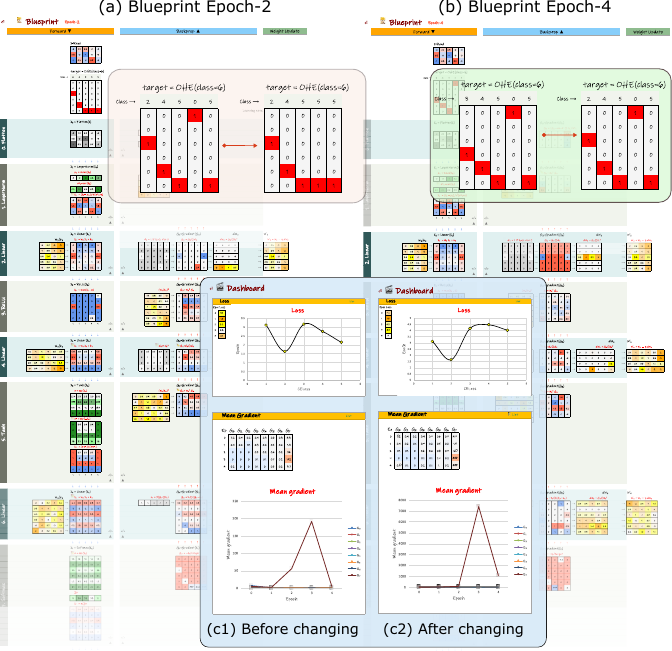}
  \caption{Learning-rate sensitivity. Changing the learning-rate cell affects downstream weight updates, epoch sheets, and dashboard summaries through workbook recalculation.}
  \label{fig:learningrate}
\end{figure}

\FloatBarrier

\subsection{Observing Vanishing Gradients}
In deeper Tanh networks, early layers can show much smaller gradient magnitudes than later layers. Because the workbook also shows the local derivative terms, users can connect the observed attenuation to the mechanism that causes it~\cite{bengio1994learning,hochreiter1998vanishing}.

\subsection{Comparing Normalization Strategies}
Layer normalization is another useful case because its internal arithmetic is often hidden inside a single library call. In \system, users can inspect the mean, variance, normalization, and downstream gradient behavior as separate worksheet regions.

\subsection{Inspecting Data Transformations}
The built-in tinyDigits sheet lets users trace how a visible sample affects the input matrix, logits, loss, and updates. This keeps the connection between the data example and the model arithmetic visible.

\subsection{Monitoring Convergence and Overfitting}
The dashboard helps users connect high-level training behavior to local arithmetic. A user can view the loss curve, then inspect whether gradients are shrinking, whether update sizes are changing, and whether weights are growing unusually large over time.

That link between summary and detail is important. Many tools provide one or the other. Here, both levels remain inside the same workbook.

\FloatBarrier

\section{Design Rationale}
\subsection{Why a Spreadsheet Representation Works}
The representation works because spreadsheets and matrix arithmetic share a structural match. Neural network computation uses arrays, repeated transformations, and explicit dependencies. Spreadsheets provide visible formulas, references, grid layout, and automatic recomputation.

An important point is that the spreadsheet does not just show a picture of the computation. The visible cells are the computation. This gives the system a form of directness that is hard to get from a separate renderer.

Prior spreadsheet-based neural-network teaching tools demonstrate the value of using worksheets for educational examples~\cite{thinYin2020simpler,semerikov2020spreadsheets}. \system differs in its emphasis on generation rather than hand-authoring. A user specifies an architecture, and the script materializes forward-pass, backward-pass, and multi-epoch training views as linked formulas. This allows the workbook to cover more than a single fixed example while preserving the inspectability of a spreadsheet artifact.

\subsection{Why the System Generates Workbooks}
One alternative would have been to hand-author a small collection of educational workbook templates. We did not choose that path because it would limit the system to fixed examples. Programmatic generation allows users to specify many different architectures while keeping the explanatory layout.

\subsection{Why Epochs Are Materialized as Separate Sheets}
Another option would have been to reuse one sheet and update it over time. Separate epoch sheets are more useful for inspection because they preserve earlier states. Users can compare snapshots, export particular epochs, and point to specific stages of training.

\subsection{Why Large Figures Matter}
The value of the system depends heavily on legibility. Large figures are not decorative in this work. They help readers verify that workbook regions, matrix blocks, gradients, and dashboard views are readable and coherent.

\section{Broader Implications}
\subsection{Productivity Software as a Research Substrate}
\system suggests a broader design pattern: instead of always building a new interface, researchers can sometimes repurpose a mature and widely used productivity tool as the interaction substrate for a new technical domain.

\subsection{Transparency as Interaction Design}
This project also treats transparency as an interface problem. It is not only about what a model predicts. It is also about how people can inspect the process that produces those predictions.

\subsection{From Explanation to Experimentation}
Because the workbook remains reactive after generation, the same artifact can act as a worked example, a debugging surface, and a small experimental sandbox.

This flexibility is one reason the system is useful as a technical report subject. The contribution is not only the static representation. It is also the way the representation supports several kinds of work with the same artifact.

\section{Discussion}
\subsection{Why Excel}
Excel introduces real trade-offs. It offers less layout control than a custom visualization system, and it was not designed for machine learning. We still chose it because it supports the design goals unusually well.

A custom table-and-tab interface could reproduce some surface features of \system, such as editable cells, tabs, and reactive updates. The advantage of Excel is that these are not isolated interface widgets. They are part of a mature computational environment that many users already understand. Users can inspect a formula through the formula bar, jump through named ranges, trace precedents and dependents, duplicate sheets for comparison, add charts, and use cross-sheet references without learning a new visualization tool.

This matters because \system treats existing spreadsheet affordances as inspection techniques. Formula inspection reveals the local computation behind a displayed value. Precedent tracing reveals upstream dependencies. Separate sheets preserve training states across epochs. Native charts let users create their own summaries after generation, as some students did during classroom deployment. A custom interface could implement these features, but doing so would require rebuilding much of the spreadsheet environment.

There are also costs. Some of Excel's broader interface remains unused, and the current prototype communicates the boundary between generated formula regions and editable control regions by convention rather than by strict enforcement. Sheet protection, clearer editable-region marking, and stronger validation could make this boundary more robust in future versions.

\begin{figure}[htbp]
  \centering
  \includegraphics[width=\textwidth]{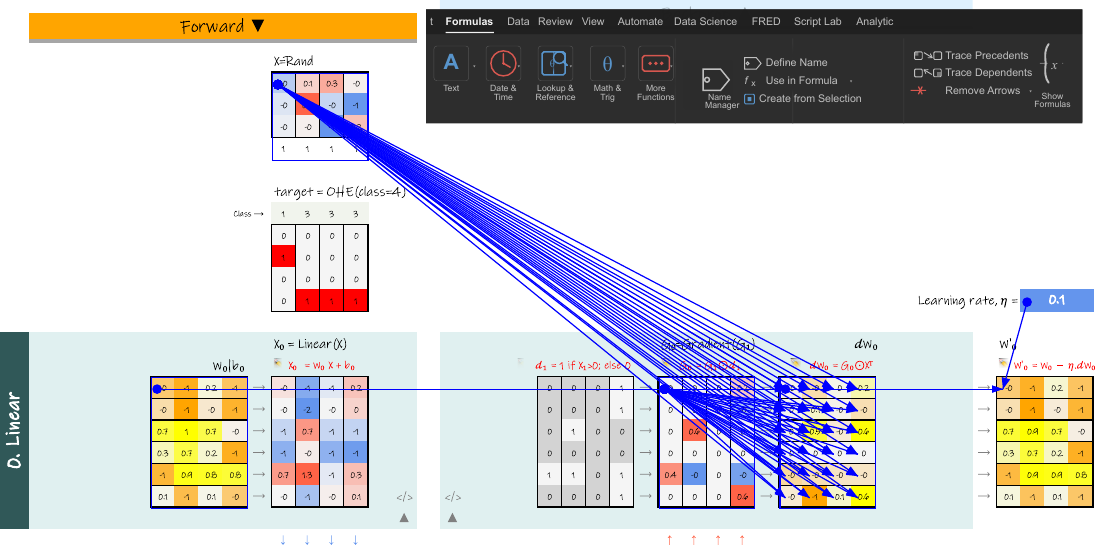}
  \caption{Using native Excel features for inspection. Precedent tracing and related tools help users follow dependencies from a selected cell to the upstream arithmetic that produced it.}
  \label{fig:tracing}
\end{figure}

\FloatBarrier

\subsection{Scope and Limitations}
The current system focuses on educational-scale and diagnostic-scale MLPs. It is most useful when matrices remain readable, so it is not meant for production-scale networks with large hidden dimensions.

The value of the approach is therefore mechanism-oriented rather than scale-oriented. It can help users understand shapes, matrix products, activation behavior, loss gradients, parameter updates, and gradient attenuation in small networks. Those mechanisms transfer conceptually to larger models, but the workbook itself does not make large production models legible. The approach fails when the relevant matrix regions become too large to inspect, when recalculation becomes slow, or when the architecture depends on non-sequential structure that the current layout cannot express compactly.

The specification language currently targets sequential architectures. Residual connections, attention, and convolutional operators would require additional layout and rendering rules. The system also relies on the bundled \texttt{tinyDigits} data and simple input modes for its included examples. Other datasets would need to be formatted into worksheet ranges with compatible dimensions and labels before the generator could route them through the blueprint.

The evidence in this report is also limited. We present representative scenarios and classroom deployment observations, but we do not yet provide a controlled study of learning outcomes, transfer, usability, or debugging performance.

There are practical limitations as well. The system depends on recent Excel features such as Office Scripts and dynamic arrays. Workbook size and recalculation cost grow with model size, layout size, and epoch count.

Even with these limitations, the prototype is already useful for the class of problems it targets. The goal is not to replace standard training pipelines. The goal is to make important computations inspectable when inspectability matters most.

\subsection{Future Directions}
Several next steps seem promising:
\begin{itemize}[leftmargin=*]
\item extending the code bridge into fuller PyTorch export,
\item adding support for convolutions, residual paths, and attention,
\item studying learning and debugging outcomes empirically, and
\item exploring similar ideas in other spreadsheet environments.
\end{itemize}

\section{Conclusion}
\system explores spreadsheets as a medium for making neural network mathematics visible, inspectable, and editable. By generating Excel blueprints that expose forward passes, losses, gradients, and updates through native formula cells and editable controls, the system helps users interact with the arithmetic of learning directly.

The system is not a replacement for mainstream machine learning tooling. Instead, it is a complementary environment for understanding, teaching, and debugging how small educational and diagnostic neural networks compute. We hope this work encourages further use of familiar productivity software as a substrate for technical transparency in AI systems.

\FloatBarrier

\appendix
\section{Additional Generated Examples}

\begin{figure*}[htbp]
  \centering
  \includegraphics[width=\textwidth]{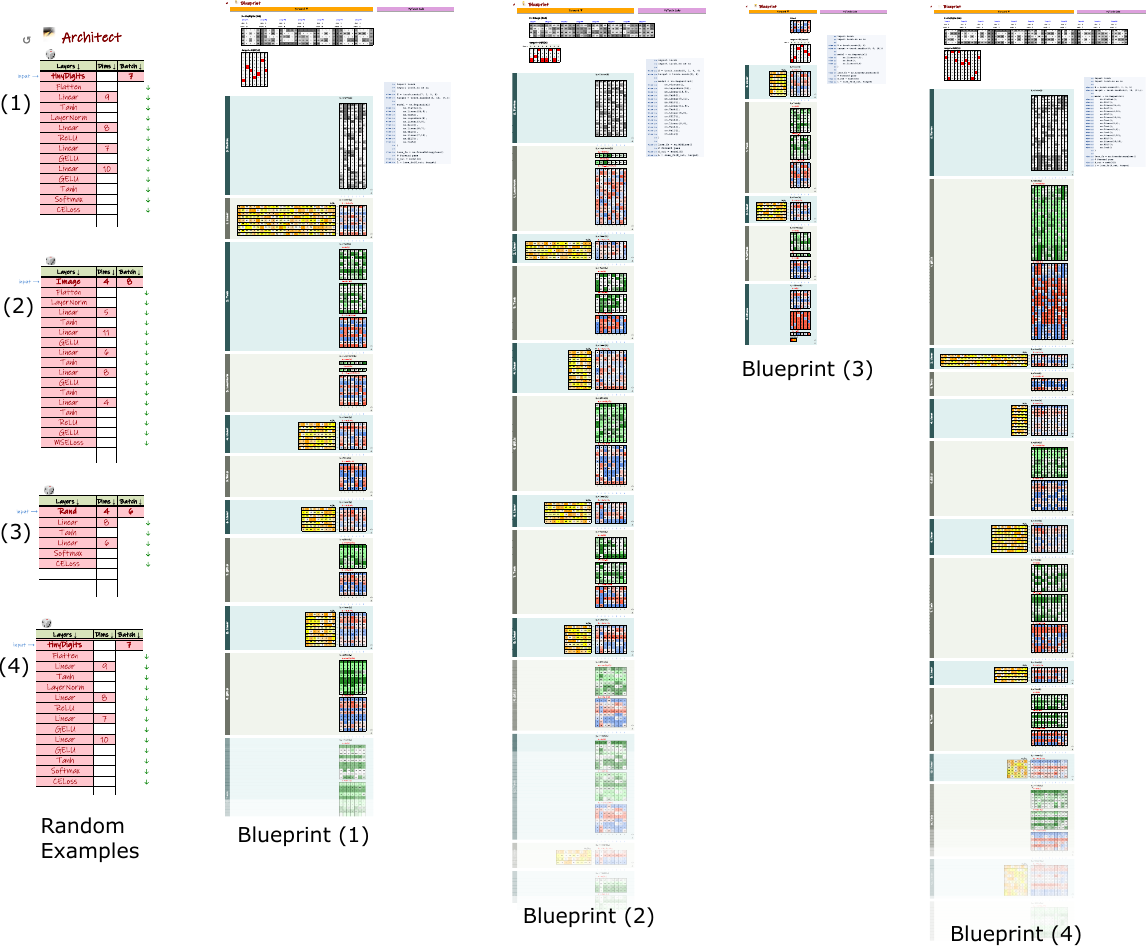}
  \caption{Random generated examples and their corresponding blueprints. The system can synthesize multiple network configurations and render their workbook views automatically.}
  \label{fig:examples}
\end{figure*}

\section{Supplementary Examples}
This appendix presents a larger collection of exported workbook examples using the same technical report style as the main document. Instead of embedding the original supplementary PDF as a separate document, we include the underlying example pages directly and organize them as appendix cases.

These appendix examples complement the main report because they show complete Architect tabs, Blueprint tabs, epoch tabs, and dashboard tabs for multiple configurations. Together, they illustrate the range of network specifications that \system can generate and the kinds of investigations the workbook supports.

The appendix covers:
\begin{itemize}[leftmargin=*]
\item simple linear and linear-plus-ReLU networks,
\item multi-layer examples with losses and normalization,
\item image-shaped inputs and flatten operations,
\item tinyDigits classification examples,
\item randomly generated architectures with multi-epoch training traces,
\item cases where users manually modify values or class indices after generation, and
\item cases where users directly edit generated formulas to investigate alternative behaviors.
\end{itemize}

For a technical report, these materials fit naturally in the appendix because they document the breadth of the system without interrupting the main narrative.

\subsection{Appendix Overview}
The examples are arranged from simple forward-only cases to larger multi-epoch training cases. Early examples focus on readability and basic operations. Later examples show random architectures, dashboards, and post-generation interventions.

\clearpage
\subsection{Example 1: Simplest Linear Architecture}
\noindent\textit{This example shows a minimal forward-only configuration with a single linear layer. It serves as the most direct illustration of the system's matrix-first representation.}
\par\medskip
\noindent\textbf{\thesubsection.1 Architect}
\begin{center}
  \includegraphics[page=1,width=0.94\textwidth]{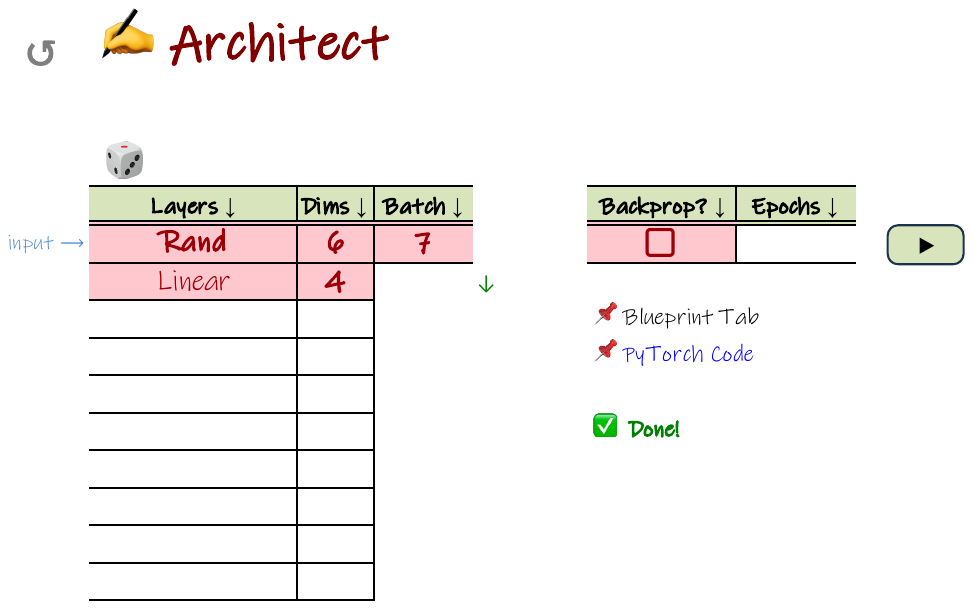}
\end{center}

\clearpage
\noindent\textbf{\thesubsection.2 Blueprint}
\begin{center}
  \includegraphics[page=2,width=0.94\textwidth]{Supps/1_basic.pdf}
\end{center}

\clearpage
\subsection{Example 2: Linear Layer Followed by ReLU}
\noindent\textit{This example extends the minimal case with a ReLU activation. It shows how the blueprint expands when a pointwise nonlinearity is inserted after a linear block.}
\par\medskip
\noindent\textbf{\thesubsection.1 Architect}
\begin{center}
  \includegraphics[page=1,width=0.94\textwidth]{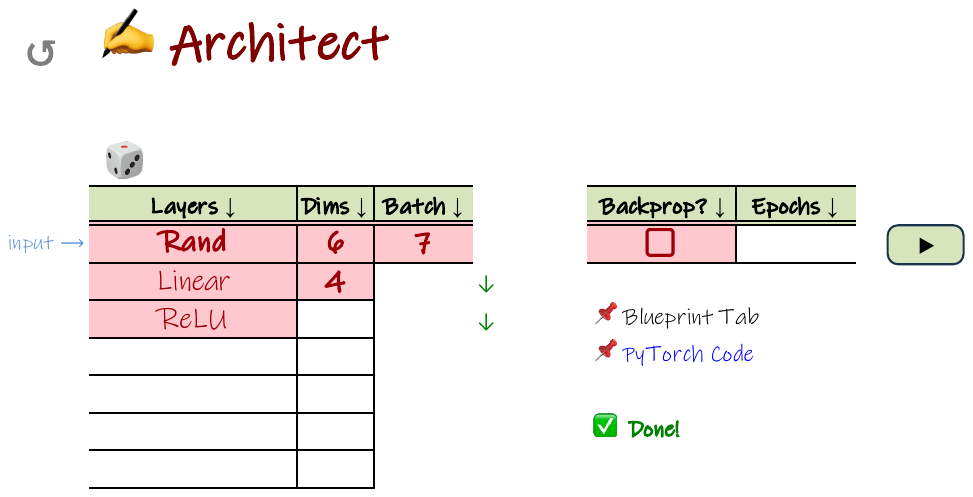}
\end{center}

\clearpage
\noindent\textbf{\thesubsection.2 Blueprint}
\begin{center}
  \includegraphics[page=2,width=0.94\textwidth]{Supps/2_relu.pdf}
\end{center}

\clearpage
\subsection{Example 3: Multi-Layer Architecture with Loss}
\noindent\textit{This example shows a deeper forward pipeline with multiple operations, including a loss function. It demonstrates how the workbook lays out a longer sequence of transformations.}
\par\medskip
\noindent\textbf{\thesubsection.1 Architect}
\begin{center}
  \includegraphics[page=1,width=0.94\textwidth]{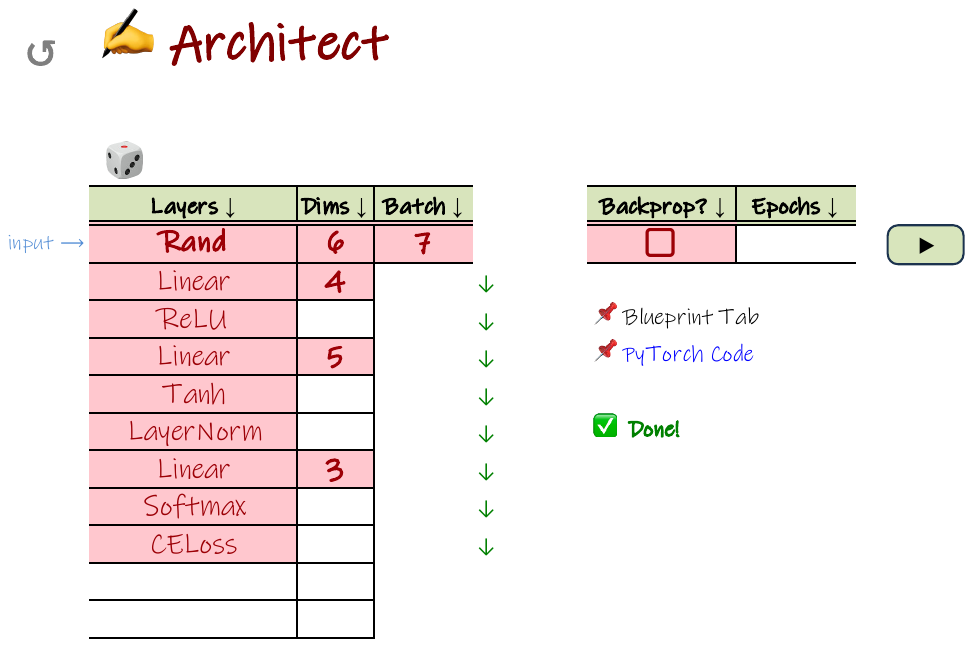}
\end{center}

\clearpage
\noindent\textbf{\thesubsection.2 Blueprint}
\begin{center}
  \includegraphics[page=2,width=0.94\textwidth]{Supps/3_all.pdf}
\end{center}

\clearpage
\subsection{Example 4: Image Input with Flatten}
\noindent\textit{This example shows the simplest image-based case. The workbook renders image-shaped input first and then flattens it into the matrix pipeline.}
\par\medskip
\noindent\textbf{\thesubsection.1 Architect}
\begin{center}
  \includegraphics[page=1,width=0.94\textwidth]{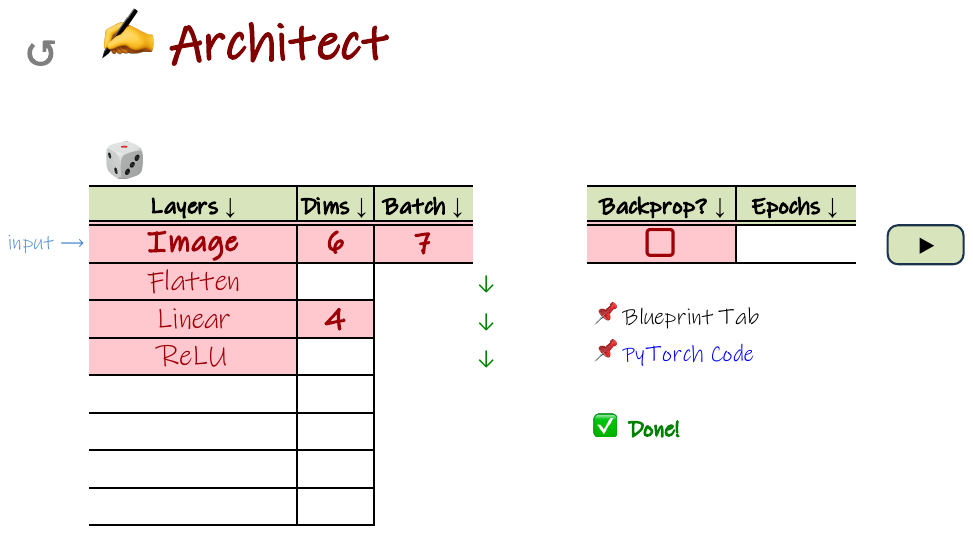}
\end{center}

\clearpage
\noindent\textbf{\thesubsection.2 Blueprint}
\begin{center}
  \includegraphics[page=2,width=0.94\textwidth]{Supps/4_Image.pdf}
\end{center}

\clearpage
\subsection{Example 5: Image Input with Additional Operations}
\noindent\textit{This example extends the image-input workflow with more layers and transformations. It shows how the same blueprint style carries over from vector inputs to image-like inputs.}
\par\medskip
\noindent\textbf{\thesubsection.1 Architect}
\begin{center}
  \includegraphics[page=1,width=0.94\textwidth]{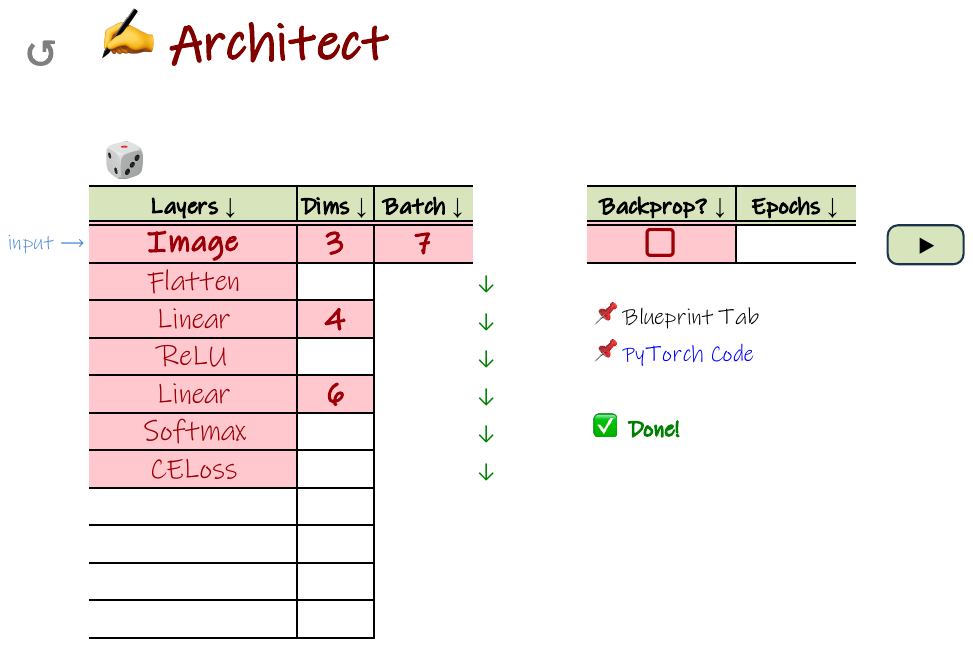}
\end{center}

\clearpage
\noindent\textbf{\thesubsection.2 Blueprint}
\begin{center}
  \includegraphics[page=2,width=0.94\textwidth]{Supps/5_Image_2.pdf}
\end{center}

\clearpage
\subsection{Example 6: tinyDigits Input}
\noindent\textit{This example uses samples from the bundled tinyDigits dataset. It demonstrates how visible dataset examples connect to the downstream network computation.}
\par\medskip
\noindent\textbf{\thesubsection.1 Architect}
\begin{center}
  \includegraphics[page=1,width=0.94\textwidth]{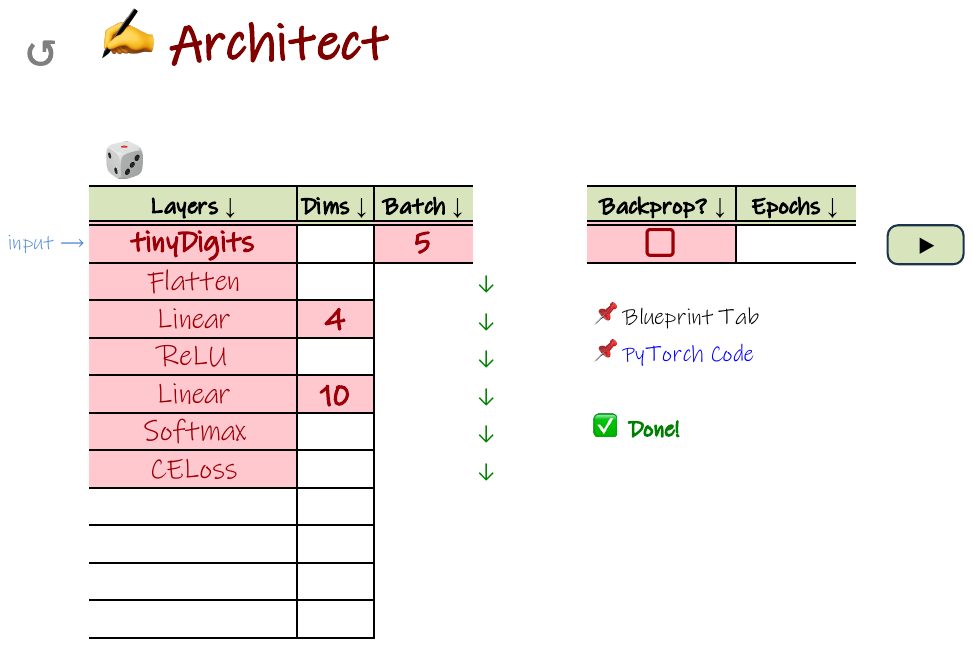}
\end{center}

\clearpage
\noindent\textbf{\thesubsection.2 Blueprint}
\begin{center}
  \includegraphics[page=2,width=0.94\textwidth]{Supps/6_tiny.pdf}
\end{center}

\clearpage
\subsection{Example 7: Random Architecture with tinyDigits and Epoch Views}
\noindent\textit{This example shows a randomly generated architecture using tinyDigits input with backpropagation and multiple epochs. It illustrates how the workbook scales from a single blueprint to a training trace with dashboard summaries.}
\par\medskip
\noindent\textbf{\thesubsection.1 Architect}
\begin{center}
  \includegraphics[page=1,width=0.94\textwidth,height=0.70\textheight,keepaspectratio]{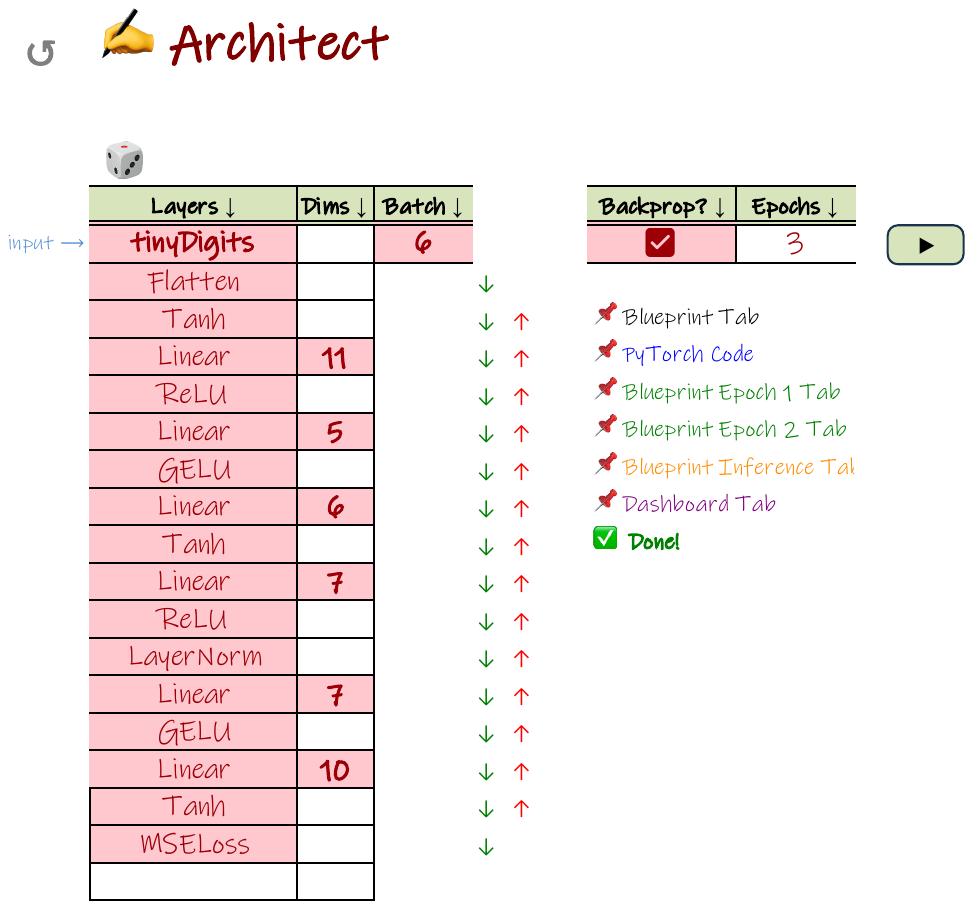}
\end{center}

\clearpage
\noindent\textbf{\thesubsection.2 Blueprint}
\begin{center}
  \includegraphics[page=2,width=0.94\textwidth]{Supps/7_epochs.pdf}
\end{center}

\clearpage
\begin{center}
  \includegraphics[page=3,width=0.94\textwidth,height=0.92\textheight,keepaspectratio]{Supps/7_epochs.pdf}
\end{center}

\clearpage
\begin{center}
  \includegraphics[page=4,width=0.94\textwidth,height=0.92\textheight,keepaspectratio]{Supps/7_epochs.pdf}
\end{center}

\clearpage
\begin{center}
  \includegraphics[page=5,width=0.94\textwidth,height=0.92\textheight,keepaspectratio]{Supps/7_epochs.pdf}
\end{center}

\clearpage
\subsection{Example 8: Random Architecture with Image Input and Multi-Epoch Training}
\noindent\textit{This example shows a larger randomly generated architecture using image input. It includes multiple epoch sheets and a dashboard, making it useful for inspecting training dynamics across time.}
\par\medskip
\noindent\textbf{\thesubsection.1 Architect}
\begin{center}
  \includegraphics[page=1,width=0.94\textwidth,height=0.70\textheight,keepaspectratio]{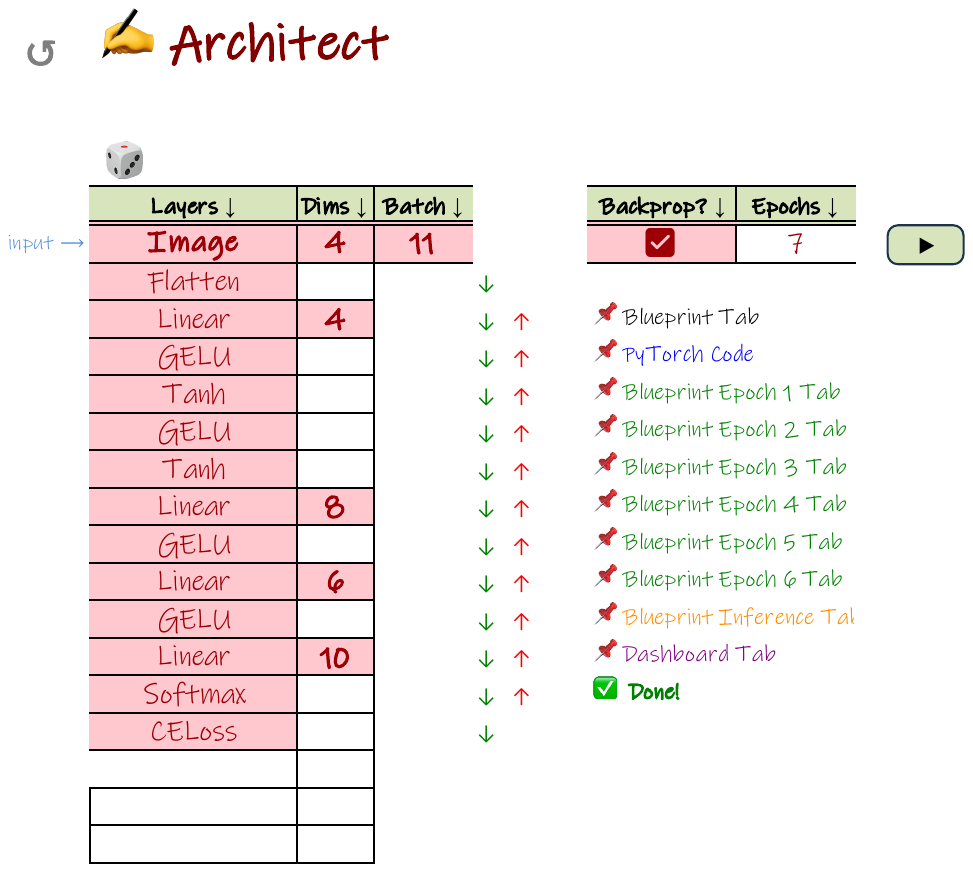}
\end{center}

\clearpage
\noindent\textbf{\thesubsection.2 Blueprint}
\begin{center}
  \includegraphics[page=2,width=0.94\textwidth]{Supps/8_epochs.pdf}
\end{center}

\clearpage
\begin{center}
  \includegraphics[page=3,width=0.94\textwidth,height=0.92\textheight,keepaspectratio]{Supps/8_epochs.pdf}
\end{center}

\clearpage
\begin{center}
  \includegraphics[page=4,width=0.94\textwidth,height=0.92\textheight,keepaspectratio]{Supps/8_epochs.pdf}
\end{center}

\clearpage
\begin{center}
  \includegraphics[page=5,width=0.94\textwidth,height=0.92\textheight,keepaspectratio]{Supps/8_epochs.pdf}
\end{center}

\clearpage
\begin{center}
  \includegraphics[page=6,width=0.94\textwidth,height=0.92\textheight,keepaspectratio]{Supps/8_epochs.pdf}
\end{center}

\clearpage
\begin{center}
  \includegraphics[page=7,width=0.94\textwidth,height=0.92\textheight,keepaspectratio]{Supps/8_epochs.pdf}
\end{center}

\clearpage
\begin{center}
  \includegraphics[page=8,width=0.94\textwidth,height=0.92\textheight,keepaspectratio]{Supps/8_epochs.pdf}
\end{center}

\clearpage
\begin{center}
  \includegraphics[page=9,width=0.94\textwidth,height=0.92\textheight,keepaspectratio]{Supps/8_epochs.pdf}
\end{center}

\clearpage
\subsection{Example 9: Post-Generation Value and Class Modifications}
\noindent\textit{This example documents a case where values and class indices are modified after generation for investigation purposes. It highlights the workbook's usefulness as an exploratory debugging surface rather than only a static export.}
\par\medskip
\noindent\textbf{\thesubsection.1 Architect}
\begin{center}
  \includegraphics[page=1,width=0.94\textwidth,height=0.70\textheight,keepaspectratio]{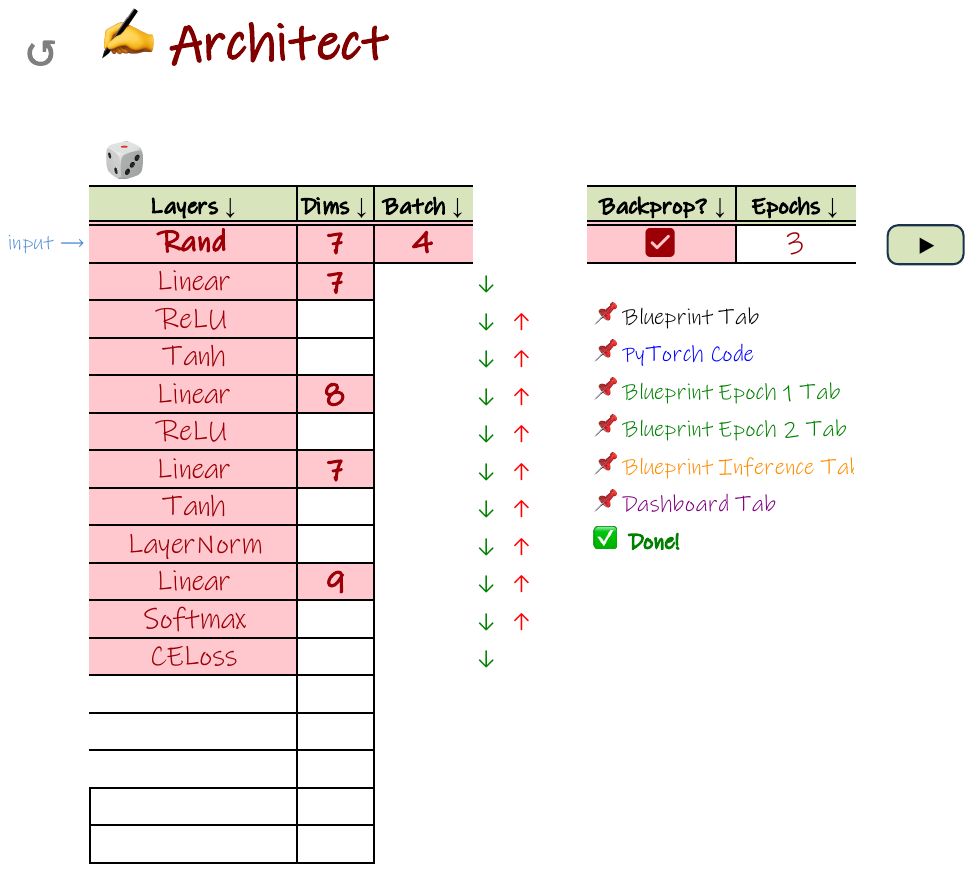}
\end{center}

\clearpage
\noindent\textbf{\thesubsection.2 Blueprint}
\begin{center}
  \includegraphics[page=2,width=0.94\textwidth]{Supps/9_play.pdf}
\end{center}

\clearpage
\begin{center}
  \includegraphics[page=3,width=0.94\textwidth,height=0.92\textheight,keepaspectratio]{Supps/9_play.pdf}
\end{center}

\clearpage
\begin{center}
  \includegraphics[page=4,width=0.94\textwidth,height=0.92\textheight,keepaspectratio]{Supps/9_play.pdf}
\end{center}

\clearpage
\begin{center}
  \includegraphics[page=5,width=0.94\textwidth,height=0.92\textheight,keepaspectratio]{Supps/9_play.pdf}
\end{center}

\clearpage
\subsection{Example 10: Post-Generation Formula Editing}
\noindent\textit{This example documents direct editing of generated Excel formulas after generation. In this case, the bias terms in the first linear layer are removed to inspect how the resulting behavior changes. This scenario highlights that the workbook remains an editable computational artifact after generation.}
\par\medskip
\noindent\textbf{\thesubsection.1 Architect}
\begin{center}
  \includegraphics[page=1,width=0.94\textwidth,height=0.70\textheight,keepaspectratio]{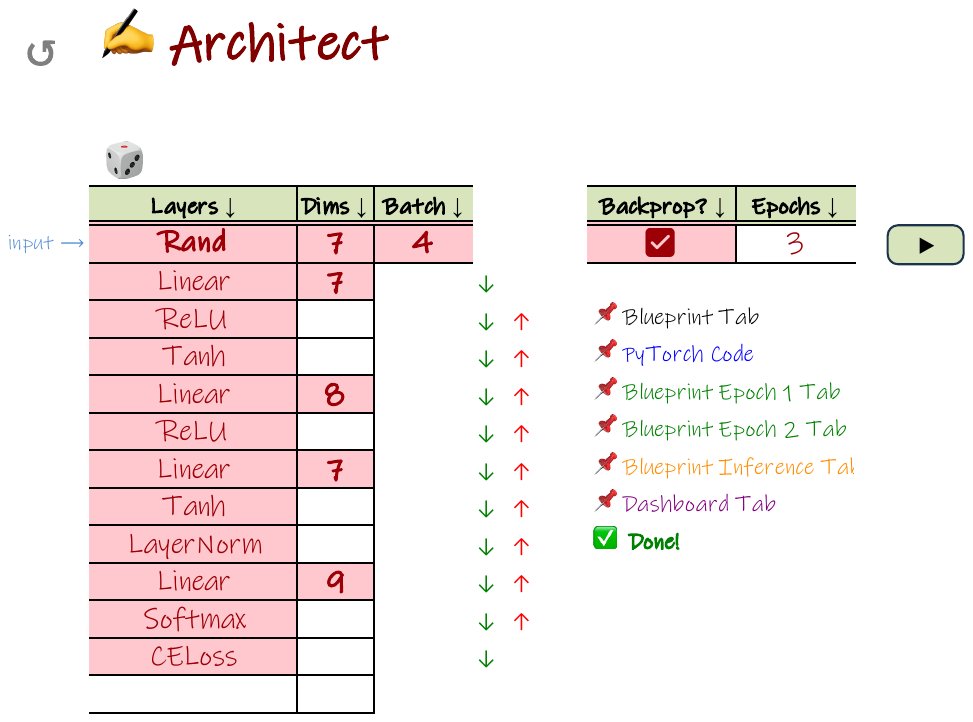}
\end{center}

\clearpage
\noindent\textbf{\thesubsection.2 Blueprint}
\begin{center}
  \includegraphics[page=2,width=0.94\textwidth]{Supps/10_noB.pdf}
\end{center}

\clearpage
\begin{center}
  \includegraphics[page=3,width=0.94\textwidth,height=0.92\textheight,keepaspectratio]{Supps/10_noB.pdf}
\end{center}

\clearpage
\begin{center}
  \includegraphics[page=4,width=0.94\textwidth,height=0.92\textheight,keepaspectratio]{Supps/10_noB.pdf}
\end{center}

\clearpage
\begin{center}
  \includegraphics[page=5,width=0.94\textwidth,height=0.92\textheight,keepaspectratio]{Supps/10_noB.pdf}
\end{center}

\bibliographystyle{plain}
\bibliography{references}

\end{document}